\documentclass[10pt]{iopart}
\usepackage{iopams}
\usepackage{graphicx}
\usepackage{bm}
\usepackage[utf8]{inputenc}
\usepackage[T1]{fontenc}
\usepackage{comment}
\usepackage{siunitx}
\usepackage{cite}
\usepackage{color}
\usepackage{cancel}

\newcommand{\vect}[1]{\ensuremath{\mathbf #1}}
\newcommand{\hatt}[1]{\ensuremath\hat{{\mathbf #1}}}

\begin{document}
\title{Experimental realization of vortex and vectorial vortex Pearcey-Gauss Beams}
\author{Valeria Rodr\'iguez-Fajardo$^1$, Gabriela Flores-Cova$^2$, Carmelo Rosales-Guzm\'an$^2$, Benjamin Perez-Garcia$^{3,*}$}
\address{$^1$Departamento de F\'isica, Universidad Nacional de Colombia Sede Bogotá,  Carrera 30 No. 45-03, Bogot\'a 111321, Colombia}
\address{$^2$Centro de Investigaciones en Óptica, A.C., Loma del Bosque 115, Colonia Lomas del Campestre, 37150 León, Gto., Mexico}
\address{$^3$Photonics and Mathematical Optics Group, Tecnologico de Monterrey, Monterrey 64849, Mexico.}
\ead{b.pegar@tec.mx}
\vspace{10pt}


\begin{abstract}
In this manuscript, we put forward two new types of structured light beams, the vortex Pearcey-Gauss (VPeG) beam, with a homogeneous polarisation distribution, and the vector vortex Pearcey-Gauss (VVPeG) beam, with a non-homogeneous polarisation distribution. The later generated as a non-separable superposition of the spatial and polarisation degrees of freedom. We also achieve their experimental realization through the combination of a spatial light modulator, which creates a scalar Pearcey-Gauss beam, and a q-plate which transforms it into a vortex or a vortex vector beam, depending on its input polarisation state. Their intensity and polarisation distribution was performed through Stokes polarimetry, along the propagation direction, which was compared with numerical simulations. As demonstrated, the VVPeG beam evolves from a pure vector beam into a vector mode of quasi-homogeneous polarisation distribution. The proposed vector beams add to the already extensive family of non-separable states of light. We anticipate that both types of beams will find applications in fields as diverse as optical metrology, optical communications, and optical tweezers, amongst others.
\end{abstract}

%
%
%
%

\section{Introduction} 
Tailoring light in their various degrees of freedom allows for the generation of novel types of beams, widely known as structured light beams, which serve as the platform for the on-demand engineering of light's properties \cite{Roadmap,Shen2022,forbes_structured_2021}. Of particular interest are structured light fields engineered as a non-separable superposition of the spatial and polarisation degrees of freedom, also known as vector beams \cite{Zhan2009,Rosales2018Review}. Such fields are giving rise to novel concepts and inspiring others, among which classical-quantum analogies have become popular \cite{forbes2019classically,Qian2017,Spreeuw1998}. In addition, structured light beams represent a powerful tool in a wide diversity of research fields such as optical communications, optical tweezers, optical metrology, and imaging, amongst many others \cite{Rosales2018Review,Yuanjietweezers2021,Ndagano2018,Milione2015,hu2019situ,BergJohansen2015,Toppel2014}. It is therefore not surprising that generating novel types of structured fields or providing novel techniques for their generation and characterisation has gained popularity in recent time. Structured beams can be classified into scalar and vectorial, the first characterized by homogeneous polarisation distributions, whereas the second by non-homogeneous \cite{Roadmap}. Both scalar and vector beams can be generated in a wide variety of ways but computer-controlled devices, such as liquid crystal Spatial Light Modulators (SLMs) or Digital Micromirror Devices (DMDs), are the choice of preference due to their high flexibility \cite{SPIEbook,Hu2022,Scholes2019,Rosales2020,Mitchell2016,Neil2009}. Nevertheless, while scalar beams are generated by manipulating only the spatial degree of freedom, vector beams require also the manipulation of polarisation in a non-separable superposition of both degrees of freedom. Crucially, even though a wide variety of structured light beams have been generated, Bessel-, Laguerre-, Ince-, Mathieu-, Parabolic-, Airy-Gauss, to mention a few \cite{rosales2021Mathieu,Galvez2012,Liyao2020,Hu2021,ZhaoBo2021,Beckley2010}, there are still many others, yet to be realized. 

Caustics and catastrophe optics are pivotal in understanding many complex light interactions.  Catastrophe optics, a branch of study within optics, explores the phenomena where light focuses to form bright patterns of illumination.  These patterns, or caustics, exhibit changes in their form, and can be classified as fold, cusp, and swallowtail, to name a few \cite{Berry1980, arnol2003catastrophe}. Examples of such caustics are Pearcey and Airy beams, both solutions to the paraxial wave equation in Cartesian coordinates. The former given in terms of the Pearcey functions \cite{Pearcey1946,Ring2012Pearcey} and the later in terms of Airy functions \cite{Efremidis2019}. On the one hand, according to catastrophe optics, Airy beams are described by the Airy function of codimension one, and their intensity pattern is underpinned by a fold caustic. Airy beams became very popular after their first experimental demonstration, in part due to the seemingly paradoxical propagation-dependent accelerating behaviour that allows them to propagate along a parabolic trajectory. On the other, Pearcey beams are patterns of codimension two and belong to the cusp caustic type, which also feature exotic properties such as auto-focusing, self-healing, and shape invariance. They have been the subject of fundamental and applied research. For example, Yihao Wang devised a method to tune in a flexible way the parameters of Pearcey beams \cite{Wang2021Pearcey}. Additional works have also reported the generation of symmetric and symmetric-odd Pearcey beams \cite{Wu2021SimPearacey,Liu2020OddPearcey}. Theoretical studies have also reported a new class of Pearcey beams that incorporate an optical vortex into its main lobe, which have been termed vortex Pearcey-Gauss (VPeG) beams \cite{LiaoSai2022,Chen_2020,Cheng2017PearceyGaussVortex}. Crucially, while various types of Pearcey beams have been generated experimentally, this is not the case for VPeG beams.

As such, in this manuscript, we propose a method to generate experimentally VPeG beams. Furthermore, we propose the experimental generation of a new class of vector beams, which we termed vector vortex Pearcey-Gauss (VVPeG) beams, a non-separable superposition of scalar VPeG beams and orthogonal polarisation states. Our approach comprises the combination of a liquid crystal SLM, where the Pearcey-Gauss beam is generated, and a q-plate\cite{Marrucci2006}, which depending on the input polarisation state can convert the Pearcey-Gauss beam into scalar VPeG or VVPeG beams. To begin with, we generate scalar VPeG beams with topological charges $\ell=\pm 1$, $\ell=\pm 2$ and performed a visual inspection of their propagation behaviour with the help of a Charge-Coupled Device (CCD) camera. Afterward, we generated two types of VVPeG beams and analysed their polarisation dynamics upon propagation. To this end, we reconstructed their polarisation distribution at various planes from $z=0$ to $z\rightarrow \infty$.

\begin{figure}[tb]
    \centering
    \includegraphics[width=0.99\textwidth]{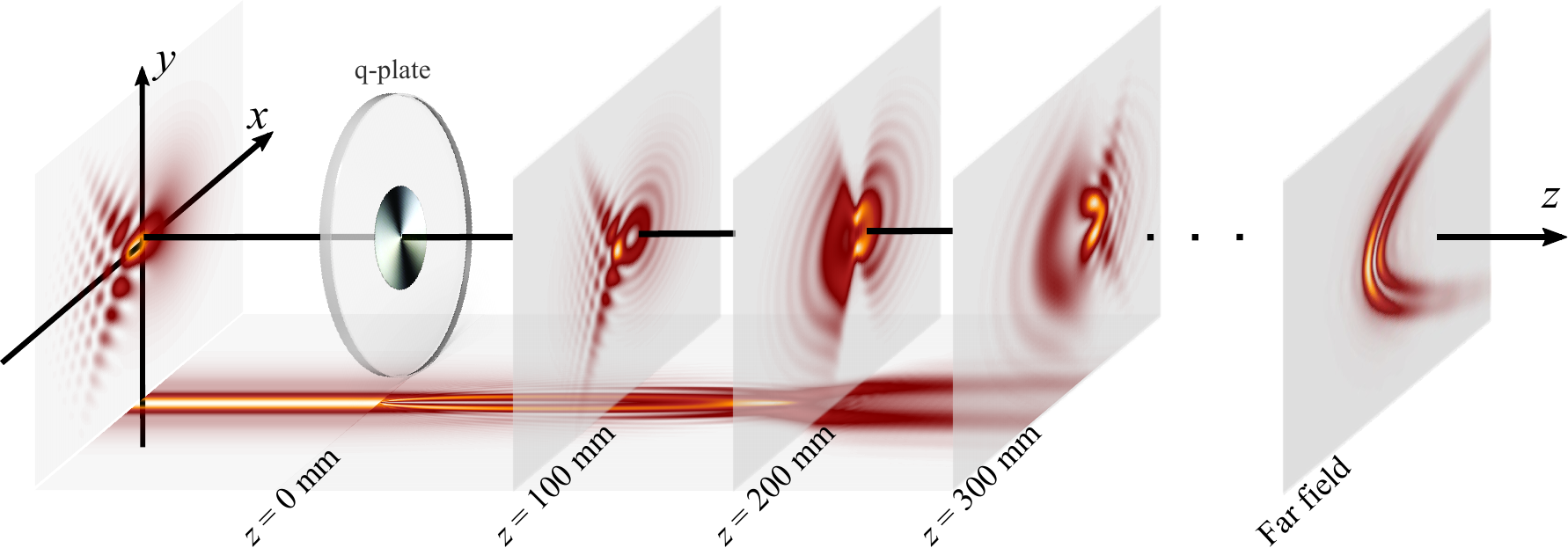}
    \caption{Schematic representation of the generation process of vector vortex Pearcey-Gauss beams. A scalar Pearcey-Gauss beam, with a transverse intensity distribution shown on the left, passes through a q-plate where it gets its vectorial nature. The intensity evolution of the beam along the propagation direction is shown at the bottom. In addition, the transverse intensity distribution is also shown at the planes $z=100, 200,\SI[]{300}{\milli\meter}$ mm and $z\rightarrow\infty$ (far-field).}
    \label{fig:concept}
\end{figure}

Figure \ref{fig:concept} shows a schematic representation of the generation process whereby a Pearcey beam, whose transverse intensity distribution is shown on the left, is sent through a q-plate, which provides the Pearcey beam with its vector nature. The intensity evolution along the propagation direction $z$ is shown at the bottom of the figure, as well as the transverse intensity distribution at the planes $z=100, 200,\SI[]{300}{\milli\meter}$ and in the far field, where the beam transforms into two parabolas. For this schematic representation, we used a $q=1/2$ q-plate.

\section{Mathematical framework} 
At the plane $z=0$, the scalar VPeG beam can be expressed as \cite{Cheng2017PearceyGaussVortex, Wu2021}
\begin{equation}\label{eq:PeG}
	\text{PeG}_\ell(x,y,0) = \exp\left(-\rho^2/w_0^2\right) \text{Pe}\left(\frac{x}{x_0 w_0},\frac{y}{y_0 w_0}\right)\exp(i\ell\varphi)\,,
\end{equation}
where $(x,y)$ and $(\rho,\varphi)$ are the Cartesian and Polar coordinates, respectively.  $w_0$ is the beam waist of the Gaussian envelope, $\ell$ is the winding number, and $x_0, y_0$ are dimensionless scaling parameters. The function $\text{Pe}(\cdot)$ is the Pearcey integral defined as \cite{Ring2012Pearcey}
\begin{equation}
	\text{Pe}(\eta,\nu) = \int_{-\infty}^{\infty} \exp\left[i(s^4 + \nu s^2 + \eta s) \right]\:ds\,.
\end{equation}
This integral can be evaluated by a suitable numerical method.  For instance, we used the so-called ``Chebyshev technology'' implemented in \cite{Driscoll2014, ApproxFun.jl-2014}.  It is important to remark that the propagation properties of scalar VPeG beams have been explored in \cite{Cheng2017PearceyGaussVortex, Wu2021}.  Here, we study the vectorial version of the VPeG beam, which is achieved by combining a $q$-plate with the generation of a PeG$_{\ell=0}$ beam via an SLM. This idea can be mathematically expressed as
\begin{equation}
\eqalign{\vect{{E}}(\vect{r}_\perp, 0) =  \text{PeG}_{\ell=0}(\vect{r}_\perp, 0) \vect{J}_q \hatt{e}_p\,,
}
\end{equation}
where $\vect{r}_\perp$ is the transverse position vector, and $\hatt{e}_p$ is the unitary polarization vector, that for convenience, might be expressed in the circular polarization basis $\{\hatt{e}_L, \hatt{e}_R\}$.  Moreover, $\vect{J}_q$ is the Jones matrix corresponding to a $q$--plate and can be written as
\begin{equation}
\eqalign{\vect{J}_q = \left[\begin{array}{cc}
				   0 & \exp(i2q\varphi)\\
				   \exp(-i2q\varphi) & 0\\
				 \end{array}\right].
}
\end{equation}
Here, $q$ is the charge of the $q$--plate, and $\varphi$ is the azimuthal angle (as previously defined).  Explicitly, we can express the VVPeG as
\begin{equation}\label{eq:vectorPGe}
	\vect{{E}}(\vect{r}_\perp, 0) = \alpha\text{PeG}_{\ell=2q}(\vect{r}_\perp, 0)\hatt{e}_L + \beta\text{PeG}_{\ell=-2q}(\vect{r}_\perp, 0)\hatt{e}_R,
\end{equation}
where $\alpha,\beta$ are complex coefficients such that $|\alpha|^2 + |\beta|^2 = 1$.  The propagated field at a given $z$--plane for each component (of Eq.\ \ref{eq:vectorPGe}) can be calculated by \cite{Goodman2005}
\begin{equation}
    E_j(\vect{r}_\perp,z)=\mathcal{F}_\perp^{-1}\{\mathcal{F}_\perp\{E_j (\vect{r}_\perp,0)\}\exp(ik_{z}(\vect{k}_\perp)z)\},
\end{equation}
where $E_j$ is the $j$--component of the vector field, $\mathcal{F}_\perp$ is the transverse Fourier transform and  $k_z(\vect{k}_\perp)$ in the paraxial approximation can be expressed as
\begin{equation}
k_z(\vect{k}_\perp) = k - \frac{|\vect{k}_\perp|^2} {2k},
\end{equation}
with $k=2\pi/\lambda$.

\section{Experimental details} 
\subsection{Experimental setup}
The experimental generation of VVPeG beams comprises the use of a Spatial Light Modulator (SLM) and a q-plate, as schematically depicted in Fig. \ref{fig:Setup}(a). The experimental setup begins with a horizontally polarised laser beam, in our case a  Helium-Neon ($\lambda=\SI[]{633}{\nano\meter}$), which is expanded and collimated with lenses $L_1$ and $L_2$ to approximate a flat wavefront. A portion of the beam, selected with a variable aperture (A), is imaged with a telescope formed by lenses $L_3$ and $L_4$ of focal length $f_3=f_4=200$ mm onto the screen of a reflective liquid crystal SLM (Holoeye Pluto 2.1 Phase Only LCOS) with a pixel resolution of $1920\times1080$ pixels and a pixel size of \SI[]{8}{\micro\meter}, where an appropriate hologram is displayed. The Pearcey-Gauss beam, which is generated in the first diffraction order, is spatially filtered with the help of a 4$-f$ system formed by lenses $L_5$ and $L_6$ of focal lengths $f_5=f_6=$ 300 mm in combination with a spatial filter located between the two lenses. This 4$-f$ system has also the purpose of imaging the Pearcey-Gauss beam onto the center of a $q-$plate, which creates the VVPeG beam. Another 4$-f$ system formed by lenses $L_7$ and $L_8$ of focal length $f_7=f_8=150$ mm relays the plane of the vector beam to the focal plane of lens $L_8$, which is precisely the plane $z=0$. The intensity of the generated beams is recorded with a charged-coupled device (CCD) camera (DCX Thorlabs, 4.65 $\mu$m pixel size), mounted on a translation stage to facilitate the scanning of the beam along the propagation direction. 

\subsection{The computer-generated hologram and the q-plate}
\begin{figure}
    \centering
    \includegraphics[width=0.99\textwidth]{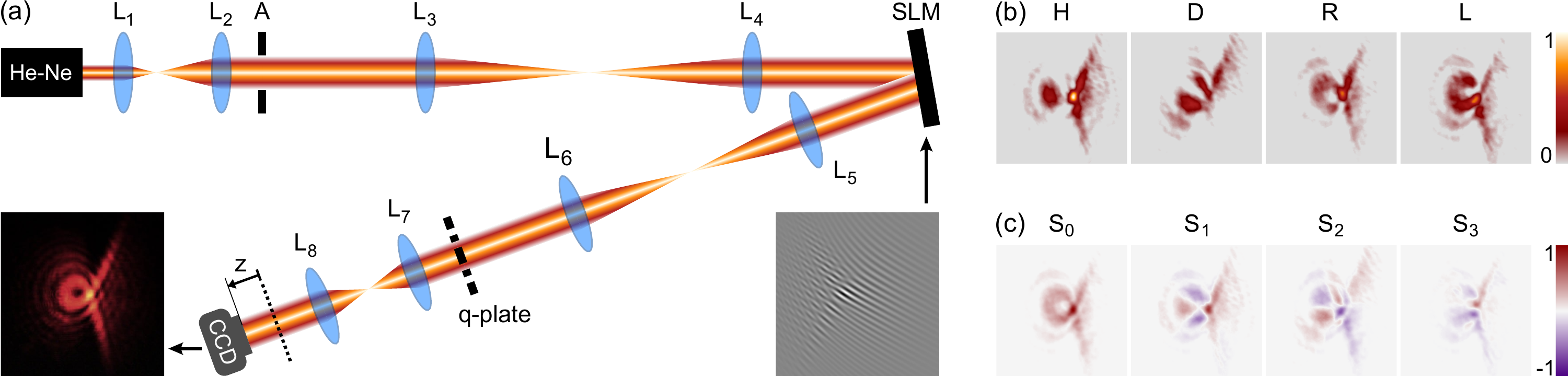}
    \caption{(a) Experimental setup for generating Scalar and vector vortex Pearcey-Gauss beams (VVPeG). Here, an expanded and collimated laser beam with horizontal polarization impinges on an SLM with the hologram (shown as an inset) to generate a Pearcey-Gauss beam is displayed. The plane of the SLM is relayed with a 4-$f$ system to coincide with the plane of a q-plate, which generates the VVPeG. Another 4-$f$ system relays the plane of the q-plate to our plane $z=0$. The intensity distribution of the beam is captured with a CCD camera mounted on a translation stage. (b) Intensity distributions of the required polarisations to determine the Stokes parameters shown in (c). For the sake of clarity, the phase retarders used to measure these intensities are not shown here.}
    \label{fig:Setup}
\end{figure}

In regards to the generation of the holograms displayed on the SLM, we first calculated numerically the complex field of the scalar Pearcey-Gauss Beam given by Eq. \ref{eq:PeG} for the specific parameters $w_0=\SI[]{1.6}{\milli\meter}$, $x_0=y_0=8\times10^{-5}$ and topological charge $\ell=0$. Further, the Pearcey function was calculated using the open-source software system ``Chebyshev technology'' implemented in MATLAB \cite{Driscoll2014}. The hologram was then created following the Arriz\'on Type 3 complex amplitude modulation method \cite{Arrizon2003} including a sinusoidal grating to spatially separate the beam of interest from the unmodulated light (zeroth order). For the sake of clarity, the central part of such a hologram is shown in the bottom-right inset in Fig.~\ref{fig:Setup}(a). The generated Pearcey-Gauss beam is then transmitted through a q-plate, which depending on the polarisation of the input beam, generates a scalar VPeG beam or a VVPeG beam. The first is achieved with right- or left-handed circular polarisation whereas the latter with linear polarisation. More precisely, under the presence of a Pearcey-Gauss beam with right-handed circular polarisation, the beam generated from the q-plate will be transformed into a scalar VPeG of topological charge $-\ell$ and opposite and orthogonal polarisation \cite{Marrucci2006}. On the contrary, if the input polarisation is left-handed, the output polarisation will be right-handed and its topological charge will be $\ell$. Crucially, if the input polarisation is a superposition of both circular polarisation states {\it i.e.} linear, the generated beam will be vectorial. Here, $q$ is a characteristic parameter of q-plates, which accounts for the number of cyclic rotations ($q$) around the center of the q-plate and is related to the topological charge by $\ell=2q$. The VVPeG beams generated from the q-plate were analyzed in propagation via Stokes Polarimetry, as explained below.

\subsection{Polarisation reconstruction via Stokes polarimetry}
The reconstruction of the transverse polarisation distribution of the generated beams was performed at various planes along the propagation direction using Stokes polarimetry. This is an intensity-based technique that requires a minimum of four intensity measurements even though it is customary to use six. The intensity measurements are used to compute a set of four parameters known as Stokes parameters, from which the polarisation distribution can be determined \cite{Goldstein2011}. The intensities and the Stokes parameters are related through the expressions,
\begin{equation}\label{Eq.SimplyStokes}
\eqalign{
 &S_{0} = I_{R}+I_{L},  \hspace{19mm} S_{1} = 2I_{H}-S_{0}\,,\\
 &S_{2} = 2I_{D}-S_{0}, \hspace{19mm} S_{3} = 2I_{R}-S_{0}\,,
}
\end{equation}
where $I_{R}$ and $I_{L}$ are the intensities that correspond to the right- and left-handed polarisation components, respectively. In addition, $I_{H}$ and $I_{D}$ are the intensities that correspond to the horizontal and diagonal polarisation components, respectively. Experimentally, $I_{R}$ and $I_{L}$ are acquired by passing the beam through the combination of a Quarter Wave-Plate (QWP) and a linear polariser (LP). To measure the first, the QWP is set to an angle of $ 45^\circ$, while the LP is to an angle of $0^\circ$, to measure the second the LP is rotated to  $90^\circ$. Finally, $I_{H}$ and $I_{D}$ are obtained by passing the beam through an LP orientated at $0^\circ$ and $45^\circ$, respectively. By way of example, Fig. \ref{fig:Setup}(b) shows a set of experimentally-measured intensities, from which the Stokes parameters shown in Fig. \ref{fig:Setup}(c) were computed. The optics to perform Stokes polarimetry were inserted between lens $L_8$ and the CCD camera but are not shown in Fig.\ref{fig:Setup}.

\section{Results and discussion} 
In the laboratory, we create VVPeG beams by generating a scalar VPeG beam using a liquid-crystal spatial light modulator and a q-plate to transform it into a vector beam. The latter modifies the phase profile and state of polarization of the incoming beam by introducing a geometrical phase \cite{Marrucci2006}. Our approach is similar to the one used by Zhou \textit{et al.} to create Airy vector beams\cite{Zhou2015}.  In addition, we simulate the system by creating a scalar horizontally linearly-polarized Pearcy beam, decomposing it into its right- and left-circular polarization components, and directly superimposing conjugated spiral phases on them. The optical vortex is placed not on the main intensity lobe as in \cite{Sun2010,Sun2011}, but in the direction the latter moves upon propagation. We propagate each component using the first Rayleigh–Sommerfeld diffraction solution and compute the far-field by applying the Fourier transforming properties of a lens \cite{Goodman2005}. We first study individual scalar VPeG beams, and then the polarization distribution of VVPeG beams, both at the far-field and as they propagate in the near-field. 
\begin{figure}
    \centering
    \includegraphics[width=0.99\textwidth]{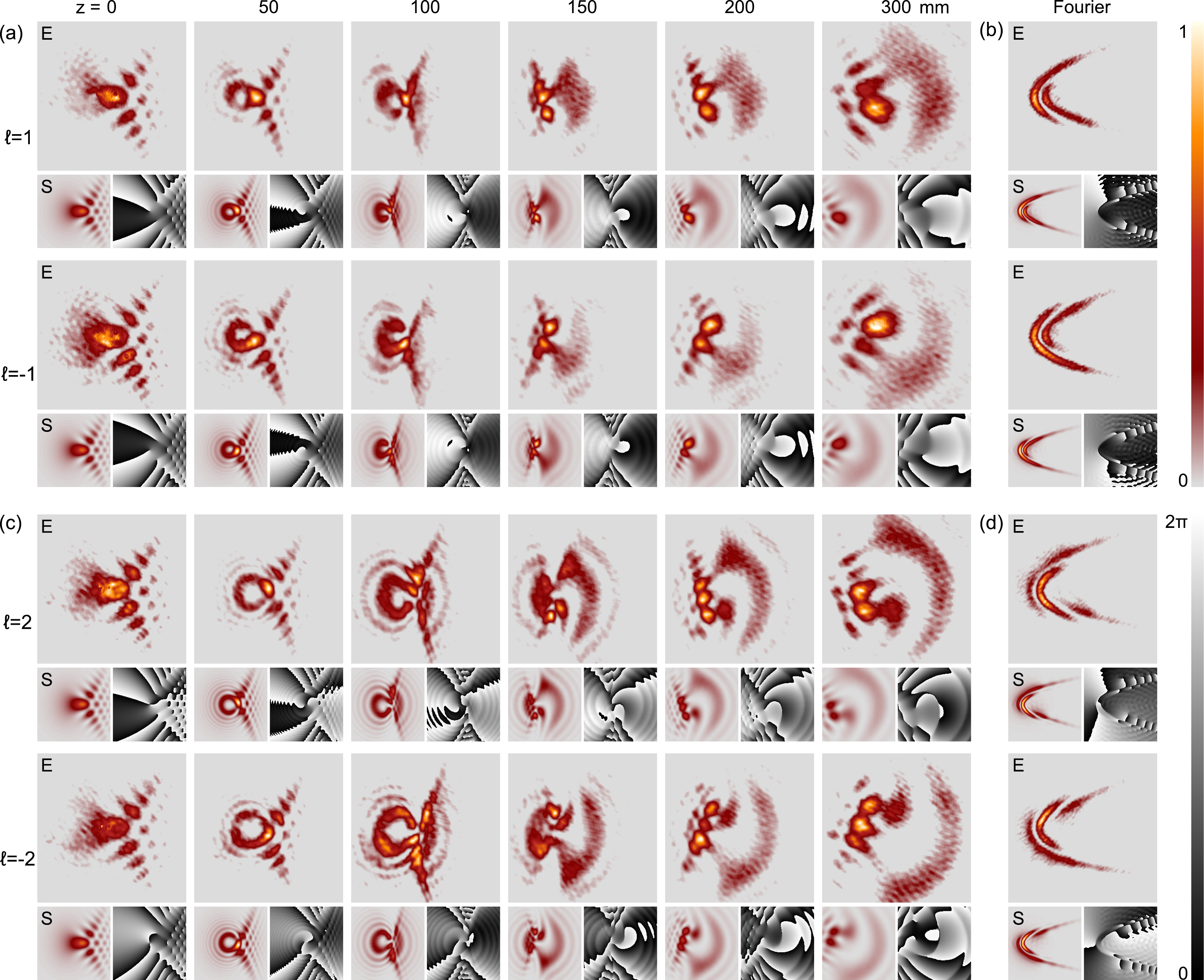}
    \caption{Transverse intensity profile of scalar vortex Pearcey-Gauss beams as a function of propagation at various planes. (a) shows the results for $\ell=1$ (top) and $\ell=-1$ (bottom), with the corresponding far-field (Fourier) distribution shown in (b). Similarly, (c) shows the results for $\ell=2$ and $\ell=-2$, top and bottom, respectively, with the far-field distribution shown in (d). Each of the experimental images is accompanied by their numerically simulated counterparts, labeled as E and S, respectively, as well as with the corresponding theoretical phase.}
    \label{fig:scalar}
\end{figure}

Figure ~\ref{fig:scalar}a presents the experimental and simulated vortex right- and left-circularly polarized scalar Pearcey-Gauss beams with topological charges $\ell=\pm1$, respectively. We generated these employing a q-plate with $q=1/2$ and took intensity images upon propagation from $z=\SI[]{0}{\milli\meter}$ to $z=\SI[]{300}{\milli\meter}$. Experimental intensity images are individually normalized to maximum intensity and simulated ones share the same normalization constant per case. While Pearcey-Gauss beams with $\ell=0$ are characterized by an invariant intensity pattern (except for asymmetrical scaling factors), this no longer holds in this case. Initially, the beam intensity pattern resembles the usual Pearcey-Gauss beam, then, at planes $z=\SI[]{50}{\milli\meter},\SI[]{100}{\milli\meter}$, the pattern features a dark core at the position of the optical vortex. This is expected since the intensity must vanish where the phase is not well defined, such that the main intensity lobe seems to circle the vortex. Thereafter, for planes $z\geq\SI[]{150}{\milli\meter}$, the intensity gradually shifts from mainly left- to right-accumulated. 
A similar behavior was observed when a VPeG beam was propagated through a harmonic potential \cite{Wu2021}, a similarity that we attribute to the fact that this kind of potential can be connected to a quadratic phase factor, which naturally appears in the Fresnel diffraction integral for propagation. Interestingly, the patterns for the $\ell=\pm1$ are vertical mirror images of one another, highlighting the fact the vortex handedness influences the pattern's shape. Upon propagation, the optical vortex moves within the beam at a different rate than the main intensity lobe does in a non-trivial manner, as can be observed in the phase images of the simulated beams. Corresponding far-field experimental intensity image and simulated field are shown in Fig.~\ref{fig:scalar}(b). A split parabola is seen as a result of the presence of the optical vortex, which forces the intensity to vanish at its position. The pattern features two half-parabolas with comparable intensities and opening to opposite sides. The $\ell=\pm1$ patterns are vertical mirrors of one another as before.

Propagated beams for topological charges $\ell=\pm2$ are presented in Fig.~\ref{fig:scalar}(c), for which we use a q-plate with $q=1$. As for $\ell=\pm1$, the intensity pattern changes upon propagation, and the $\ell=\pm2$ beams are mirror images of one another. In this case, however, the dark core at the center is larger, as expected due to the optical vortex density limitation \cite{Roux2003}. Far-field experimental intensity images and simulated field are shown in Fig.~\ref{fig:scalar}(d). The parabola is split into three branches, the central one being of higher intensity than the other two, due to the initial optical vortex $\ell=\pm2$ separating into two vortices of $\ell=\pm1$, as can be seen in the simulated field's phase. Interestingly, this separation occurs early in the propagation, as evinced in the image corresponding to $z=\SI[]{50}{\milli\meter}$. A similar splitting was theoretically observed for the far-field by Cheng \textit{et al.} \cite{Cheng2017}.

\begin{figure}
    \centering
    \includegraphics[width=0.99\textwidth]{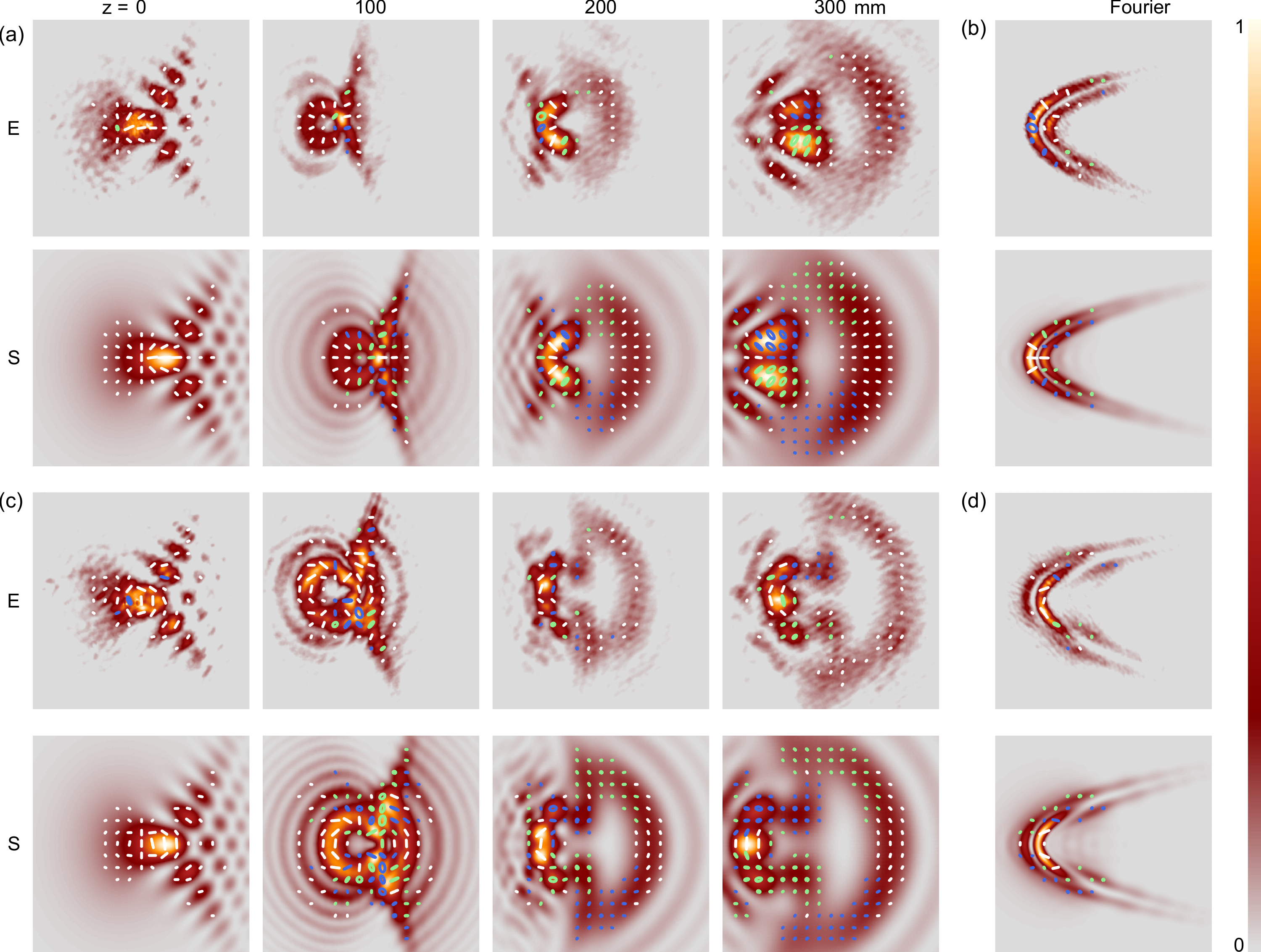}
    \caption{Transverse polarisation distribution overlapped with the intensity profile of vector vortex Pearcey (VVPeG) beams at various planes along the propagation direction. (a) and (b) correspond to VVPeG of topological charges $\ell=\pm 1$ and $\ell=\pm 2$, respectively, experiment (E) on top and numerical simulations (S) on the bottom. (c) and (d) correspond to the results in the far-field. Further, green and blue ellipses represent right- and left-handed circular polarisation, whereas white lines represent linear polarisation.}
    \label{fig:vector}
\end{figure}

We generated two examples of VVPeG beams, described by Eq.~\ref{eq:vectorPGe} with $\alpha=\beta= 1/\sqrt{2}$ for $q=1/2$ and $q=1$. As for the scalar beams, we studied the beams in propagation and at the far-field (Fourier plane of a lens). Fig~\ref{fig:vector} presents polarization distributions overlapped over the intensity normalized profiles for the two cases, each with the experimental and simulated results in the top and bottom rows, respectively. In this figure, green and blue ellipses are associated to the right- and left-handed circular polarisation, whereas the white lines to the linear. The intensity profiles, contrary to the scalar case, are symmetric about the horizontal axis. This is expected since the total intensity of the vector beam is the sum of the individual intensities of its polarization constituents, that being vertical mirror images of one another, complement each other to form the symmetric pattern. In the first case (Fig~\ref{fig:vector}a), the polarization distribution resembles radial polarization up to $z=\SI[]{100}{\milli\meter}$, with some of them slightly elliptical or slanted from the radial direction. While initially, most polarization states are linear, upon propagation other states appear, such that at $z=\SI[]{300}{\milli\meter}$ it approximates a Full Poincar\'e beam \cite{Brown2010}. The far-field beam features two parabolas with comparable maximum intensities that merge into a single one away from the center. A similar intensity pattern was theoretically obtained for a scalar Pearcy-Gauss vortex beam with $\ell=-1$ by Cheng \textit{et al.} \cite{Cheng2017}. In the second case (Fig~\ref{fig:vector}b), two parabolas are observed with two additional streaks pointing to the center of the brightest parabola. Here, the polarization distribution evolves from spider-like at $z=\SI[]{0}{\milli\meter}$ to a richer structure with multiple states of polarization at $z=\SI[]{300}{\milli\meter}$. 

\section{Conclusions} 
Complex vector light beams have reshaped the landscape of modern optics, not only due to their exotic properties but also to the wide variety of applications they are enabling. In this manuscript, we implement two new families of beams, vortex Pearcey-Gauss (VPeG) and vector vortex Pearcey-Gauss (VVPeG) beams, which are based on the well-known Pearcey-Gauss beams, and, to the best of our knowledge, are the first to produce them experimentally. We first generated a scalar Pearcey-Gauss beam, using a liquid crystal spatial light modulator, and transmitted this through a q-plate to obtain a vector beam. 
By way of example, we generated two types of scalar and vector vortex Pearcey-Gauss beams with two different q-plates $q=1/2$ and $q=1$ and analysed their propagation behaviour. In the scalar case, as the beams propagate from the near- to the far-field, the intensity distribution evolves from a Pearcey-Gauss beam with a vortex near its main lobe, to a parabolic structure formed by two or more semi-parabolas (depending on the topological charge). Such behaviour deserves a detailed analysis but it is out of the scope of this manuscript and therefore it is left for a future article. In the vectorial case, the intensity profile evolves similarly to the previous case, with the main difference that far-field features complete parabolas. A more interesting behaviour occurs in the polarisation distribution, which evolves from completely mixed states into quasi-scalar states, in a similar way as in\cite{Hu2021}. For completeness, we compared our experimental results with numerical simulations showing a good agreement. Even though here we only presented experimental results for VVPeG beam with equal weights $\alpha=\beta$, the idea can be easily extended to non-equal weights by using additional phase retarders, as in\cite{cardano2012}. Furthermore, even though for simplicity we generated VVPeG beams with the combination of an SLM and a q-plate, Eq. \ref{eq:PeG} allows to generate VPeG beams directly from an SLM, which will enable the generation of other VVPeG beams by interferometric means. Finally, we anticipate that the new types of scalar and vector beams presented here will also inherit the self-healing property of the Pearcey-Gauss beam \cite{Ring2012Pearcey,Kovalev2015}.

\section*{References}

\begin{thebibliography}{10}
\expandafter\ifx\csname url\endcsname\relax
  \def\url#1{{\tt #1}}\fi
\expandafter\ifx\csname urlprefix\endcsname\relax\def\urlprefix{URL }\fi
\providecommand{\eprint}[2][]{\url{#2}}

\bibitem{Roadmap}
Rubinsztein-Dunlop H, Forbes A, Berry M~V, Dennis M~R, Andrews D~L, Mansuripur M, Denz C, Alpmann C, Banzer P, Bauer T, Karimi E, Marrucci L, Padgett M, Ritsch-Marte M, Litchinitser N~M, Bigelow N~P, Rosales-Guzm{\'{a}}n C, Belmonte A, Torres J~P, Neely T~W, Baker M, Gordon R, Stilgoe A~B, Romero J, White A~G, Fickler R, Willner A~E, Xie G, McMorran B and Weiner A~M 2017 {\em J. Opt.\/} {\bf 19} 013001

\bibitem{Shen2022}
Shen Y and Rosales-Guzmán C 2022 {\em Laser \& Photonics Reviews\/} {\bf 16} 2100533

\bibitem{forbes_structured_2021}
Forbes A, de~Oliveira M and Dennis M~R 2021 {\em Nat. Photon.\/} {\bf 15} 253--262

\bibitem{Zhan2009}
Zhan Q 2009 {\em Adv. Opt. Photonics\/} {\bf 1} 1--57

\bibitem{Rosales2018Review}
Rosales-Guzm\'{a}n C, Ndagano B and Forbes A 2018 {\em J. Opt.\/} {\bf 20} 123001

\bibitem{forbes2019classically}
Forbes A, Aiello A and Ndagano B 2019 Classically entangled light {\em Progress in Optics\/} (Elsevier Ltd.) pp 99--153

\bibitem{Qian2017}
Qian X~F, Vamivakas A~N and Eberly J~H 2017 {\em Opt. Photon. News\/} {\bf 28} 34--41

\bibitem{Spreeuw1998}
Spreeuw R~J~C 1998 {\em Found. Phys.\/} {\bf 28} 361--374

\bibitem{Yuanjietweezers2021}
Yang Y, Ren Y, Chen M, Arita Y and Rosales-Guzmán C 2021 {\em Adv. Photonics\/} {\bf 3}

\bibitem{Ndagano2018}
Ndagano B, Nape I, Cox M~A, Rosales-Guzm\'{a}n C and Forbes A 2018 {\em J. Light. Technol.\/} {\bf 36} 292--301

\bibitem{Milione2015}
Milione G, Lavery M~P, Huang H, Ren Y, Xie G, Nguyen T~A, Karimi E, Marrucci L, Nolan D~A, Alfano R~R {\em et~al.\/} 2015 {\em Opt. Lett.\/} {\bf 40} 1980--1983

\bibitem{hu2019situ}
Hu X~B, Zhao B, Zhu Z~H, Gao W and Rosales-Guzm{\'a}n C 2019 {\em Optics Letters\/} {\bf 44} 3070--3073

\bibitem{BergJohansen2015}
Berg-Johansen S, T\"{o}ppel F, Stiller B, Banzer P, Ornigotti M, Giacobino E, Leuchs G, Aiello A and Marquardt C 2015 {\em Optica\/} {\bf 2} 864--868

\bibitem{Toppel2014}
T{\"o}ppel F, Aiello A, Marquardt C, Giacobino E and Leuchs G 2014 {\em New J. Phys.\/} {\bf 16} 073019

\bibitem{SPIEbook}
Rosales-Guzm\'{a}n C and Forbes A 2017 {\em How to shape light with spatial light modulators\/} (SPIE Press)

\bibitem{Hu2022}
Hu X~B and Rosales-Guzm{\'{a}}n C 2022 {\em Journal of Optics\/} {\bf 24} 034001

\bibitem{Scholes2019}
Scholes S, Kara R, Pinnell J, Rodríguez-Fajardo V and Forbes A 2019 {\em Optical Engineering\/} {\bf 59} 1 -- 12

\bibitem{Rosales2020}
Rosales-Guzm{\'a}n C, Hu X~B, Selyem A, Moreno-Acosta P, Franke-Arnold S, Ramos-Garcia R and Forbes A 2020 {\em Scientific Reports\/} {\bf 10} 10434

\bibitem{Mitchell2016}
Mitchell K~J, Turtaev S, Padgett M~J, \v{C}i\v{z}m\'{a}r T and Phillips D~B 2016 {\em Opt. Express\/} {\bf 24} 29269--29282

\bibitem{Neil2009}
Savage N 2009 {\em Nature Photonics\/} {\bf 3} 170--172

\bibitem{rosales2021Mathieu}
Rosales-Guzm{\'a}n C, Hu X~B, Rodr{\'\i}guez-Fajardo V, Hernandez-Aranda R~I, Forbes A and Perez-Garcia B 2021 {\em Journal of Optics\/} {\bf 23} 034004

\bibitem{Galvez2012}
Galvez E~J, Khadka S, Schubert W~H and Nomoto S 2012 {\em Appl. Opt.\/} {\bf 51} 2925--2934

\bibitem{Liyao2020}
Yao-Li, Hu X~B, Perez-Garcia B, Bo-Zhao, Gao W, Zhu Z~H and Rosales-Guzm{\'a}n C 2020 {\em Applied Physics Letters\/} {\bf 116} 221105

\bibitem{Hu2021}
Hu X~B, Perez-Garcia B, Rodr\'{i}guez-Fajardo V, Hernandez-Aranda R~I, Forbes A and Rosales-Guzm\'{a}n C 2021 {\em Photon. Res.\/} {\bf 9} 439--445

\bibitem{ZhaoBo2021}
Zhao B, Rodríguez-Fajardo V, Hu X~B, Hernandez-Aranda R~I, Perez-Garcia B and Rosales-Guzmán C 2021 {\em Nanophotonics\/}  20210255

\bibitem{Beckley2010}
Beckley A~M, Brown T~G and Alonso M~A 2010 {\em Opt. Express\/} {\bf 18} 10777--10785

\bibitem{Berry1980}
Berry M~V and Upstill C 1980 {\em Progress in Optics\/} {\bf 18}(C) 257--346

\bibitem{arnol2003catastrophe}
Arnol'd V, Wassermann G and Thomas R 2003 {\em Catastrophe Theory\/} (Springer Berlin Heidelberg) ISBN 9783540548119 \urlprefix\url{https://books.google.com.mx/books?id=GQoQyqia45gC}

\bibitem{Pearcey1946}
Pearcey T 1946 {\em The London, Edinburgh, and Dublin Philosophical Magazine and Journal of Science\/} {\bf 37}(268) 311--317

\bibitem{Ring2012Pearcey}
Ring J~D, Lindberg J, Mourka A, Mazilu M, Dholakia K and Dennis M~R 2012 {\em Opt. Express\/} {\bf 20} 18955--18966

\bibitem{Efremidis2019}
Efremidis N~K, Chen Z, Segev M and Christodoulides D~N 2019 {\em Optica\/} {\bf 6} 686--701

\bibitem{Wang2021Pearcey}
Wang Y 2021 {\em J. Opt. Soc. Am. A\/} {\bf 38} 1726--1731

\bibitem{Wu2021SimPearacey}
Wu Y, Zhao J, Lin Z, Huang H, Xu C, Liu Y, Chen K, Fu X, Qiu H, Liu H, Wang G, Yang X, Deng D and Shui L 2021 {\em Opt. Lett.\/} {\bf 46} 2461--2464

\bibitem{Liu2020OddPearcey}
Liu Y, Xu C, Lin Z, Wu Y, Wu Y, Wu L and Deng D 2020 {\em Opt. Lett.\/} {\bf 45} 2957--2960

\bibitem{LiaoSai2022}
Sai L, Ke C, Hong-wei H, Ceng-hao Y, Meng-ting L and Wang-xuan S 2023 {\em Chinese Optics\/} {\bf 16} 1195--1205

\bibitem{Chen_2020}
Chen X, Xu C, Yang Q, Luo Z, Li X and Deng D 2020 {\em Chinese Physics B\/} {\bf 29} 064202

\bibitem{Cheng2017PearceyGaussVortex}
Cheng K, Lu G and Zhong X 2017 {\em Applied Physics B\/} {\bf 123} 60 ISSN 1432-0649

\bibitem{Marrucci2006}
Marrucci L, Manzo C and Paparo D 2006 {\em Phys. Rev. Lett.\/} {\bf 96} 163905

\bibitem{Wu2021}
Wu J, Xu C, Wu L and Deng D 2021 {\em Optics Communications\/} {\bf 478} 126367

\bibitem{Driscoll2014}
Driscoll T~A, Hale N and Trefethen L~N 2014 {\em Chebfun Guide\/} (Pafnuty Publications)

\bibitem{ApproxFun.jl-2014}
Olver S and Townsend A 2014 A practical framework for infinite-dimensional linear algebra {\em Proceedings of the 1st Workshop for High Performance Technical Computing in Dynamic Languages -- HPTCDL `14\/} ({IEEE})

\bibitem{Goodman2005}
Goodman J~W 2005 {\em Introduction to Fourier optics\/} (Roberts \& company)

\bibitem{Arrizon2003}
Arriz\'{o}n V 2003 {\em Opt. Lett.\/} {\bf 28} 1359--1361

\bibitem{Goldstein2011}
Goldstein D~H 2011 {\em Polarized light\/} (CRC Press)

\bibitem{Zhou2015}
Zhou J, Liu Y, Ke Y, Luo H and Wen S 2015 {\em Opt. Lett.\/} {\bf 40} 3193--3196

\bibitem{Sun2010}
Sun X~W, Liu Y~J, Luo D and Dai H~T 2010 {\em Optics Letters, Vol. 35, Issue 23, pp. 4075-4077\/} {\bf 35}(23) 4075--4077 ISSN 1539-4794

\bibitem{Sun2011}
Sun X~W, Liu Y~J, Luo D and Dai H~T 2011 {\em Optics Letters, Vol. 36, Issue 9, pp. 1617-1619\/} {\bf 36}(9) 1617--1619 ISSN 1539-4794

\bibitem{Roux2003}
Roux F~S 2003 {\em Optics Communications\/} {\bf 223}(1-3) 31--37 ISSN 0030-4018

\bibitem{Cheng2017}
Cheng K, Lu G and Zhong X 2017 {\em Optik\/} {\bf 149} 189--197

\bibitem{Brown2010}
Brown T~G, Alonso M~A and Beckley A~M 2010 {\em Optics Express, Vol. 18, Issue 10, pp. 10777-10785\/} {\bf 18}(10) 10777--10785 ISSN 1094-4087

\bibitem{cardano2012}
Cardano F, Karimi E, Slussarenko S, Marrucci L, de~Lisio C and Santamato E 2012 {\em Applied Optics\/} {\bf 51}(10) C1 ISSN 1559-128X \urlprefix\url{https://opg.optica.org/abstract.cfm?URI=ao-51-10-C1}

\bibitem{Kovalev2015}
Kovalev A~A, Kotlyar V~V, Zaskanov S~G and Porfirev A~P 2015 {\em Journal of Optics\/} {\bf 17}(3) 035604

\end{thebibliography}
\providecommand{\newblock}{}

\end{document}